\documentclass[prd,showpacs,amsmath,showkeys,twocolumn,floatfix]{revtex4} 
\usepackage{epsfig,dcolumn}
\usepackage{graphicx}
\usepackage{graphicx}
\usepackage{bm}

\usepackage[usenames]{color}

\def\xb{{x_B}}   
\def\wtilp{{{\widetilde p}}}
\def\wtilk{{{\widetilde k}}}
\def\wtilH{{\widetilde{H}^+}}
\def\fqx{{f_q(x)}}
\def\bfqx{{\bar{f}_q(x)}}
\def\epUD{{\epsilon^{\mu\nu}_{\ \ 03}}}

\def\qs{\displaystyle{\not} q }
\def\ks{\displaystyle{\not} k }
\def\ps{\displaystyle{\not} p}

\def\lsim{\mathrel{\rlap{\lower4pt\hbox{\hskip1pt$\sim$}}
    \raise1pt\hbox{$<$}}}
\def\gsim{\mathrel{\rlap{\lower4pt\hbox{\hskip1pt$\sim$}}
    \raise1pt\hbox{$>$}}}
\begin{document}


\title{ Regge Exchange Contribution to Deeply Virtual Compton Scattering}

\author{ Adam P. Szczepaniak and J.T.~Londergan}
\affiliation{ Physics Department and Nuclear Theory Center \\
Indiana University, Bloomington, Indiana 47405 }

\author{ Felipe J. Llanes-Estrada } 
\affiliation{Departamento de F$\acute{\imath}$sica Te$\acute{o}$rica I, \\
Universidad Complutense de Madrid, 28040 Madrid, Spain } 
\date{\today}

\begin{abstract}
Recently we have shown that exclusive QCD photon-induced reactions 
at low Mandelstam-$t$ are best described by Regge exchanges in the entire 
scaling region, and not only for low values of Bjorken-$x$. 
In this paper we explore this crucial Regge behavior in Deeply Virtual 
Compton Scattering from the point of view of collinear factorization, with 
the proton tensor written in terms of Generalized Parton Distributions, 
and we reproduce  this feature. Thus it appears that in the Bjorken limit,
a large class of hard, low-$t$ exclusive processes are more sensitive to 
the meson cloud of the proton than to its fundamental quark structure.
These process will then be described most efficiently by 
process-dependent Regge Exclusive Amplitudes rather than by universal 
Generalized Parton Distributions. We introduce such Regge Exclusive Amplitudes 
for Deeply Virtual Compton Scattering.
\end{abstract}


\maketitle
\section{Introduction\label{Sec:Intro}}

Nowadays it is common to consider deeply virtual exclusive electroproduction 
of mesons or photons in the context of generalized parton distributions 
(GPD's) \cite{Ji:1998pc, Radyushkin:1997ki, Goeke:2001tz, Belitsky:2001ns, Diehl:2003ny}. 
In the case of deeply virtual Compton scattering 
(DVCS) \cite{Ji:1996nm,Radyushkin:1996nd}, a single initial quark near 
its mass shell becomes highly virtual after it interacts with the off-shell 
photon. This  virtual quark propagates essentially without interaction with 
the quark and gluon spectators until it radiates a real photon.  To produce a 
final hadron in place of a real photon, the off-shell quark can radiate a hard 
gluon that enhances its correlation with soft quarks and antiquarks around the 
target, and hence increases the probability of hadronization into a single 
meson.
 
In the language of QCD factorization \cite{Collins:1998be,Collins:1996fb}, 
exchange of a single off-shell quark between the interaction point for the 
virtual photon and real photon (or meson, in the case of hard exclusive meson 
production) corresponds to 
application of the operator product expansion to a product of electromagnetic 
currents and/or meson interpolating fields. In the center of mass frame  
between the initial quark and virtual photon, the incoming quark has one large 
momentum component and is nearly on its mass shell.  To obtain amplitudes in 
terms of generalized parton distributions it is necessary to ignore the other, 
small momentum components and quark masses.

An alternative approach to exclusive processes involving strong interactions 
is based on 
identifying singularities of the scattering amplitude close to the physical 
region of the relevant kinematical variables. In particular one expects 
physical states with quantum numbers of the $t$-channel to dominate processes 
involving two-to-two particle scattering, $ab \to cd$ at low momentum transfer 
to the target, $t=(p_d - p_b)^2<0$  and large values of $s$ (the square of 
the center of mass energy) $s = (p_a + p_b)^2>> -t$. Inclusion of all allowed 
$t$-channel exchanges leads to the Regge-type  dependence, 
of the scattering amplitude $A(s,t)$ 
\begin{equation}
A(s,t) \sim s^{\alpha(t)}
\label{eq:Astdef}
\end{equation}  
on the center of mass energy $s$ where for low momentum transfer 
$-t   \alt  1\mbox{ GeV}^2$. In Eq.~(\ref{eq:Astdef}) the intercept 
$\alpha(t)$ of the Regge trajectory is a positive 
number less than one (with exception of diffractive scattering which in this language corresponds to the Pomeron exchange).  The microscopic picture of exclusive electroproduction 
suggested by Regge phenomenology differs substantially from an explanation 
predicated on generalized parton distributions. Qualitatively, the Regge 
picture corresponds to virtual photon scattering from quarks in the meson 
cloud around the bare nucleon, as opposed to the virtual photon scattering 
from quarks in the core of the nucleon, as implied by the connection between 
GPD's and the spatial distribution of valence quarks \cite{Bur03,Ji97}.  

A distinguishing feature of the GPD mechanism 
is amplitude scaling in terms of Bjorken variables, {\it i.e.} at fixed 
momentum transfer and mass of the produced hadron (or photon) the hadronic 
part of electroproduction amplitudes \cite{Vanderhaeghen:1999xj} is predicted 
to be a function of $-q^2 = Q^2$, the photon virtuality and 
$\xb \equiv Q^2/(2\nu)$ where $\nu = p_a\cdot p_b$ is the energy of the 
virtual photon in the target rest frame.  Furthermore at fixed Bjorken $\xb$ 
QCD makes specific predictions for the leading order large-$Q^2$  dependence. 
In the Regge case, the amplitude is expected to be a function of both $Q^2$ 
and $\nu$. 

Recent results on exclusive vector and pseudoscalar meson production from 
JLab \cite{Morand:2005ex} and HERMES \cite{HERMES-from-jlab} 
generally do not exhibit the $Q^2$ scaling predicted by QCD. In the case of 
meson production, the $\gamma^* p \to M p$ cross-section is predicted to fall 
as $1/(Q^2)^n$ with $n=3$, while Jlab $\omega$ production data and HERMES 
$\pi^+$ data taken in a similar kinematic range give $n \sim 2$. 
Earlier data \cite{Airapetian:2000ni} on $\rho_0$ production might be
consistent with QCD expectations, but these results appear to be 
softer than the $n=3$ predicted by QCD scaling. 

DVCS data from Hall A at Jefferson Laboratory 
\cite{MunozCamacho:2006hx} and HERMES \cite{HERMES-from-jlab} appear to be 
consistent with the 
$Q^2$-independent amplitude predicted by QCD \cite{Vanderhaeghen:1998uc}, 
however the available $Q^2$ window is quite small, from 
$1.5 - 2.5\,\mbox{GeV}^2$ and within the published experimental errors one 
cannot rule out a power-like dependence of the amplitude, $A \propto 
(Q^2)^\alpha$, with $\alpha$ as large as 0.25.  Perhaps even more surprising, 
"standard" Regge-exchange models have proved successful in describing a 
variety of differential cross sections \cite{Morand:2005ex,Kubarovsky}, in the 
kinematical range where scaling would be expected based on comparisons with 
deep inelastic scattering (DIS).  As we see it, a fundamental question is 
whether the success of the Regge picture is accidental. If not, this 
immediately raises the question of how one can disentangle scattering off 
the meson cloud from effects of nucleon tomography. 

It is well known that Regge exchanges also contribute to DIS 
structure functions \cite{Donnachie:1998gm}, but their contribution is 
restricted to very low $\xb \sim 0$. Once it was realized that Regge 
exchange may play a significant role in exclusive electroproduction, attempts 
have been made to incorporate Regge effects using analogies with DIS, 
{\it i.e} to restrict Regge contributions in exclusive electroproduction 
reactions to low-$\xb$ so that scaling is not otherwise modified 
\cite{Ahmad:2006gn,Guzey:2005ec,Kumericki:2007sa,Jenkovszky:2006bq}. 
To the best of our knowledge it has not been 
proven that Regge contributions should only contribute to exclusive amplitudes 
in this domain, and in fact in Ref.~\cite{Szczepaniak:2006is} we provide 
arguments that, in a certain kinematic regime, Regge effects should be 
substantial even at large $\xb$.

In this paper, we investigate further this issue.  In 
Ref.~\cite{Szczepaniak:2006is} we analyzed hard exclusive reactions by 
examining the high-energy behavior of $t$-channel exchange processes.  
Here we will show that utilization of an $s$-channel 
framework, in which one analyzes the ``hand-bag'' diagrams that are used 
in extracting GPDs, leads to the same conclusions reached in 
Ref.~\cite{Szczepaniak:2006is}, \textit{i.e.} we show that in the region 
of high energy and small $t$, Regge effects should make sizeable 
contributions to hard exclusive amplitudes. 

In this $s$-channel formalism we are able to explore the interplay between 
Regge behavior in the 
parton-nucleon amplitude and the hard interaction induced by the virtual 
photon. We will show that there are crucial differences between DIS and DVCS 
hand-bag diagrams which make Regge components of the soft parton-nucleon 
amplitude much more pronounced for DVCS than for DIS. We find that 
the difference between these processes arises when one attempts a 
collinear factorization of the quark propagators occuring in these 
processes. In the presence of Regge behavior in the parton-nucleon 
amplitude, the DVCS formalism is ill-defined. We then compute the 
hand-bag diagram using the full hard quark propagator. For hard 
exclusive processes, the divergent terms that are introduced produce 
a non-analytic, non-scaling dependence on the photon virtuality.  

This has the following effect on the hard exclusive amplitudes.  First, 
the breakdown of factorization means that the soft amplitudes are not 
universal, but are process-dependent.  Second, in the region of small 
$t$ Regge effects will make substantial contributions to DVCS and 
exclusive meson production.  Third, the $Q^2$ behavior of these 
hard exclusive processes should be different from that predicted from  
scaling arguments.  

Our paper is organized as follows. In the following Section we introduce 
the framework and consider the case of collinear factorization. We review 
both DIS and DVCS reactions, and we show that the DVCS formalism is 
ill-defined in the presence of Regge-like behavior in the parton-nuclon 
amplitude. In Section~\ref{Sec:DVCSwo} we compute the hand-bag diagram with 
the full hard quark propagator and show how the divergent would-be 
collinear factorization forces a non-analytical, non-scaling dependence on
the photon virtuality. We derive the $Q^2$ behavior for hard exclusive 
processes and show how it differs from scaling predictions, and how this 
$Q^2$ behavior is related to the leading Regge trajectories.  We analyze 
existing DVCS and exclusive meson data, and show that their $Q^2$ behavior 
is, at least qualitatively, consistent with our predictions.

\section{Collinear factorization in presence of Regge asymptotics
\label{Sec:One}} 

The hadronic tensor that describes electromagnetic transitions in double 
diagonal DIS $\gamma^* p \to \gamma^* p$ or off-diagonal $\gamma^* p \to 
\gamma p$ DVCS reactions, is given by 
\begin{equation}
T^{\mu\nu}  = i  \int d^4z e^{i\frac{q'+q}{2} z} \langle p'\lambda' | 
  T J^\mu(z/2) J^\nu(-z/2) | p\lambda\rangle.
\label{Wdef}
\end{equation} 
In Eq.~(\ref{Wdef}), $q$ is the four momentum of the virtual photon, 
$q^2 < 0$, 
$q' = q + p - p'  \equiv q - \Delta$, and  $q'^2 = 0$ is the momentum of 
the real photon produced in DVCS. In the case of DIS, $q'^2 = q^2$ and 
$\Delta = 0$ and the DIS cross section is proportional to the discontinuity 
of $T$ across the cut in $(p+q)^2$. Even though we will explicitly consider 
only the kinematics relevant for either DIS or DVCS the analysis can easily be 
extended to the more general case of arbitrary time-like $q'$ which is 
relevant, for example for meson electroproduction.  The currents are given by 
$J^\mu(z) = \sum_q e_q J^\mu_q(z)$, $J^\mu_q(z) =\bar{\psi}(z) \gamma^\mu 
\psi(z)$ where $\psi$ is the quark field operator and $e_q$ is the 
quark charge. Throughout this paper we will consider a single quark flavor. 
For large $Q^2$ the $z$-integral peaks at $z^2 \sim~1/Q^2$ and using the 
leading order operator product expansion of QCD we replace the product of the 
two currents by a product of two quark field operators and a free propagator 
between the photon interaction points $(z/2,-z/2)$  
 \begin{widetext} 
 \begin{eqnarray}
 T^{\mu\nu}  & =  & - e_q^2 \int \frac{d^4 z d^4k }{(2\pi)^4}     
  \frac{ \left[\gamma^\mu \left(\ks + \frac{\qs+\qs'}{2} \right)
 \gamma^\nu\right]_{\alpha\beta} }{\left( \frac{q+q'}{2} + k \right)^2 
  + i\epsilon}
 \langle p'\lambda'|T \overline{\psi}_\alpha(z/2) \psi_\beta(-z/2) 
  | p\lambda\rangle \left[ e^{-ikz} - e^{+ikz}(\mu \leftrightarrow \nu)\right] 
  \nonumber \\
 & \equiv & -i e_q^2\int \frac{d^4k}{(2\pi)^4} \left\{ 
   \frac{ \left[\gamma^\mu \left(\ks + \frac{\qs+\qs'}{2} \right)
  \gamma^\nu\right]_{\alpha\beta}}{\left( \frac{q+q'}{2} + 
  k \right)^2 + i\epsilon} - 
    \frac{ \left[\gamma^\nu \left(-\ks + \frac{\qs+\qs'}{2} \right)
  \gamma^\mu\right]_{\alpha\beta}}{\left( \frac{q+q'}{2} - 
 k \right)^2+i\epsilon} \right\} A_{\beta\alpha}(k,\Delta,p,\lambda,\lambda').
 \nonumber \\
\end{eqnarray} 
\end{widetext} 
Here $A$ is the parton-nucleon scattering amplitude un-truncated, with respect 
to the parton legs,  
\begin{eqnarray}
  A_{\beta\alpha} &\equiv& A_{\beta\alpha}(k,\Delta,p,\lambda',\lambda) 
  \nonumber \\ 
   A_{\beta\alpha} &=& 
   -i \int d^4z e^{-ikz} \langle p'\lambda'|T \bar\psi_\alpha(z/2) 
  \psi_\beta(-z/2) | p\lambda\rangle. \nonumber \\
\label{Adef} 
\end{eqnarray}
As in Refs.~\cite{Landshoff:1970ff, Landshoff:1972,Brodsky:1971zh,Brodsky:1972vv,BCG}, 
we assume that despite its non-physicality, 
the analytical properties of the parton-nucleon amplitude display structures 
in the complex plane similar to conventional hadron scattering amplitudes. 
This is necessary if such amplitudes  are to be of any use at all, {\it i.e.} 
if they are to be connected to asymptotic properties of QCD~\cite{alkofer}. 
The $T$-ordered product could 
then be replaced by a normal ordered product corresponding to generalized 
parton distributions~\cite{T}. For the purpose of our study it will be more 
efficient to deal directly with  the $T$-ordered amplitudes.   
 \begin{figure}
\includegraphics[width=4in]{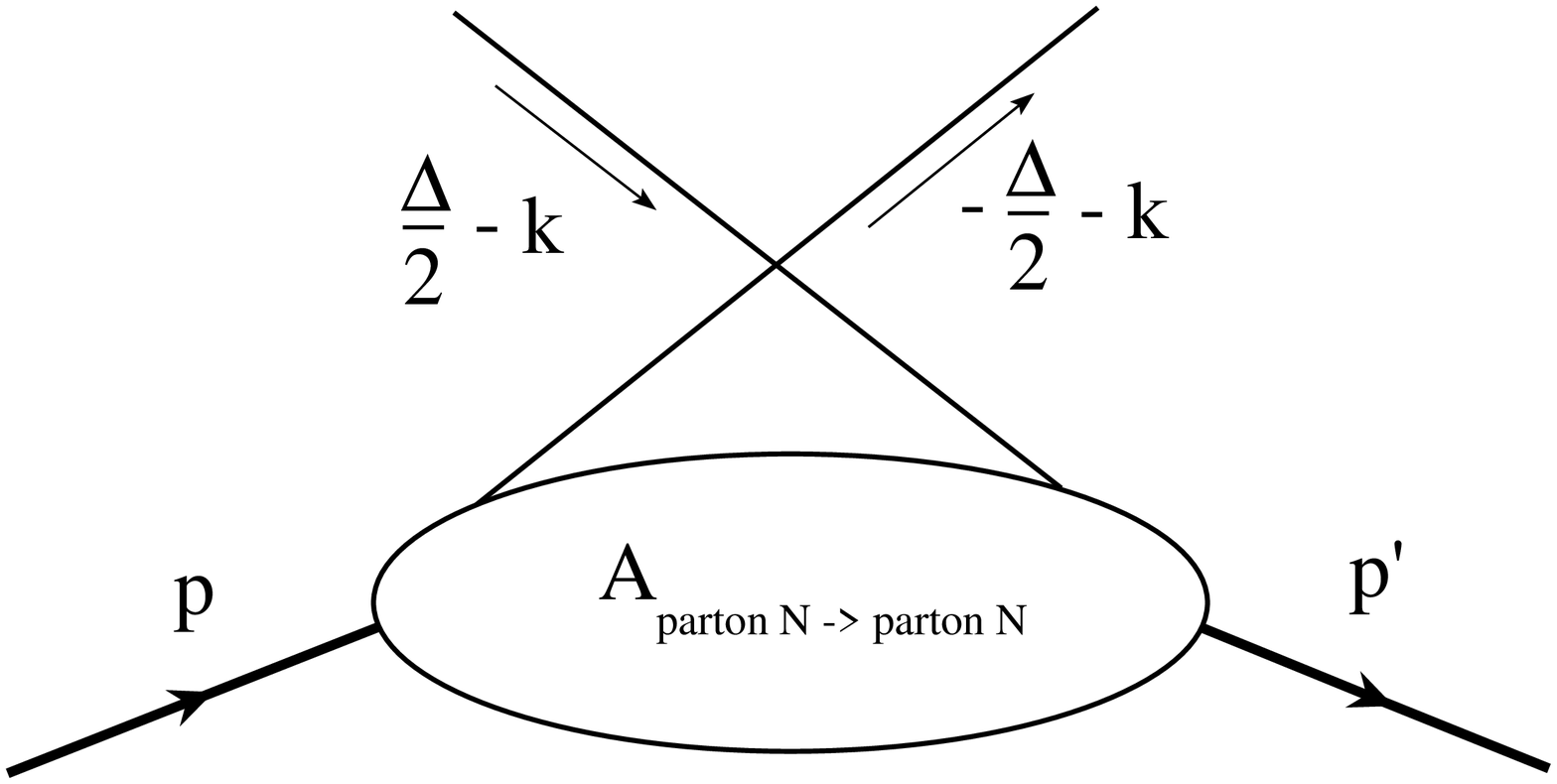}
\caption{\label{A} } Parton-nucleon scattering amplitude
\end{figure}
 The parton-nucleon amplitude is a function of four variables and the nucleon 
helicities. The variables are  $k^2_1 = 
(\Delta/2 -k)^2$, $k^2_2=(-\Delta/2-k)^2$ the (virtual) masses of the incoming 
and outgoing partons ($\Delta = p'-p=q-q'$), $s = (p+k_1)^2 = 
[ (p+p')/2 -k]^2$ is the square of the center of mass energy in the 
$s$-channel, $u=(p' - k_1)^2 
   = [(p'+p)/2 + k]^2$ is the square of the center of mass energy in the 
$u$-channel. Together with the four-momentum transfer, $t = (p' - p)^2 = 
\Delta^2$ they satisfy $s + t + u = k_1^2 + k^2_2 + 2M^2$ where $M$ is the 
nucleon mass. 
  
To obtain DIS scaling relations it is necessary to assume that the 
parton-nucleon amplitude has cuts for positive $s$ and $u$. We will be 
interested primarily in the implications of the high-$s$ or $u$ behavior at 
low $t$ where the amplitude is expected to be helicity conserving. Furthermore 
to reproduce the scaling limit of DIS and to preserve current conservation, 
the dependence on the parton spin (Dirac) indices must be of the form 
$A_{\beta\alpha} \propto \ks_{\beta\alpha}$, or $[\gamma_5 \ks]_{\beta\alpha}$ 
The former (latter) contributes respectively to the symmetric 
(antisymmetric) parts of the hadronic tensor $T^{\mu\nu}$. Finally for 
fixed-$t$ we arrive at the 
general representation of the parton-nucleon amplitude in the form, 
\begin{equation} 
    A = A^+  \frac{\ks_{\beta\alpha}}{4}  \delta_{\lambda'\lambda} + 
     A^- \frac{\left[ \gamma_5 \ks \right]_{\beta\alpha}  }{4} 
      \tau^{3}_{\lambda'\lambda} \ \ , 
\label{Aform}
\end{equation} 
with the factor of $1/4$ introduced for later convenience and with the 
amplitudes $A^{\pm}$ having  the Mandelstam representation, 
 \begin{widetext} 
\begin{equation}
A^{\pm}   =  (2\pi)^4 \int \frac{d\mu^2_1 d\mu_2^2 dm^2} 
 { (\mu_1^2 - k_1^2 - i\epsilon)(\mu^2_2 - k_2^2 - i\epsilon)} 
\left[ \frac{\rho^{\pm} _s(\mu_1^2,\mu_2^2,m^2,t) }{m^2 - s  -  i\epsilon}
   \pm  \frac{\rho^{\pm}_u(\mu_1^2,\mu_2^2,m^2,t)}{m^2 - u - i\epsilon} 
  \right]+   \mbox{subtractions}.
\label{Apmdef}
\end{equation}
\end{widetext}

At asymptotically high energies, the quark and antiquark structure functions 
are becoming identical 
 which implies that the $s$ and $u$-channel spectral functions  
become identical, and so for large $m^2$ $\rho^\pm_u \sim \rho^\pm_s$. 
These amplitudes are in principle different in the valence (finite $m^2$) 
region. Even though we are primarily interested in the large-$m^2$ region we 
will distinguish between the $s$ and $u$ spectral functions in order to be 
able to keep track of quark and antiquark contributions. The dependence of the 
spectral density $\rho^\pm$ on $\mu_1^2$ and $\mu^2_2$ determines the 
dependence of the parton-nucleon scattering amplitude on parton virtualities. 
In perturbation theory \cite{Landshoff:1970ff} one would have  
$\rho \propto \delta(\mu_1^2 - m_q^2) \delta(\mu_2^2 - m_q^2)$
 where $m_q$ is the bare quark mass.  For the bound state nucleon, however, 
  $\rho$ is expected to be softer~\cite{Landshoff:1972,BCG}, {\it e.g}  in 
order to reproduce correctly the large $Q^2$ fall-off of the form factors 
\cite{Brodsky:2005vt}). 
The spin structure of $A$ could be more complicated than given by the two 
terms in Eq.~(\ref{Aform}), for example there could be terms proportional to 
$\ps$, $\ps'$, or $\ps\gamma_5$
  {\it etc.}~\cite{Landshoff:1972}. As will be clear from the the discussion 
that follows, however, it is the terms proportional to $\ks$ that lead to the 
Regge behavior of the structure functions and thus will be considered here. 
Without loss of generality we can take 
\begin{widetext}
\begin{equation}
\rho^\pm_{u,s}(\mu_1^2,\mu_2^2,m^2) \to    \rho^\pm_{u.s}(m^2,t) (\mu^2)^n 
  \frac{d^n}{d(\mu^2)^n} \left[  \delta(\mu_1^2 - \mu^2) 
  \delta(\mu_2^2 - \mu^2)  \right] ,
\end{equation} 
\end{widetext}
with $n \ge 1$, where for simplicity we use a single scale $\mu$ for both
partons (inclusion  of charge symmetry breaking effects is an obvious
generalization).  The most general spectral density can always be written as 
a linear combination of functions of this type $\rho=\sum_n c_n \rho_n$.
Henceforth we will omit the subindex on $\rho_n$. As we have already 
discussed, for low $m^2$ this amplitude is expected to be sensitive to poles 
and cuts associated with low energy resonances and few-particle production 
thresholds.  For large $m^2$ it is expected to be dominated by the leading 
Regge trajectory, 
\begin{equation}
 \rho^\pm_{u,s}(m^2,t) = \rho^\pm_{u,s,V}(m^2,t) + \rho^\pm_{u,s,R}(m^2,t).
\end{equation} 
For large $m^2$, the valence part $\rho^\pm_{u,s,V}(m^2)$   falls off with 
$m^2$ and does not require subtractions,  On the contrary for large $m^2$ the 
Regge part, $\rho^\pm_{u,s,R}(m^2)$ behaves as 
\begin{equation}
\rho^\pm_{u,s,R}(m^2,t) \to \beta^\pm_{u,s}(t) \left( \frac{m^2}{\mu^2}
  \right)^{\alpha^\pm_{u,s}(t) },
\end{equation} 
where $0< \alpha_{u,s}^\pm(t) < 1$ for small $t$ and requires one 
subtraction in Eq.~\ref{Apmdef}. Here we consider only the quark contribution, 
as opposed to gluon exchanges which lead to diffractive, Pomeron-type 
contributions with $\alpha>1$. These could also be effectively included but 
would require additional subtractions.  As we are interested in the 
low-$t$ limit, we have approximated the intercepts and residues by their 
values in the limit $t\to 0$.

\begin{figure}
\includegraphics[width=3.5in]{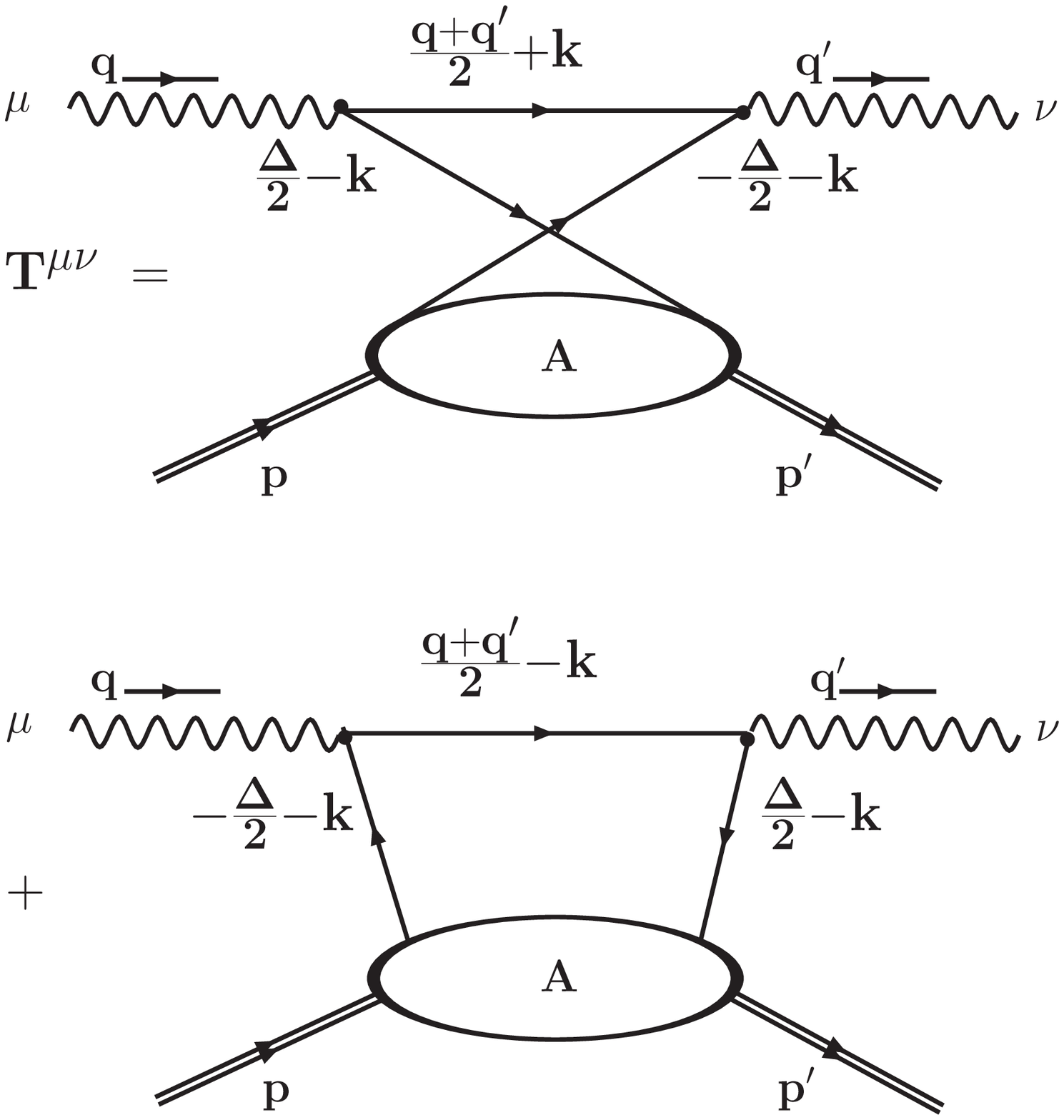}
\caption{\label{W} }  $u$ and $s$ channel contributions to the DVCS amplitude
\end{figure}

In the following we will be interested in the role of the Regge (high 
energy) component and  thus the parton-nucleon amplitude can be written, 
\begin{widetext} 
\begin{equation} 
A^\pm (s,u,k_1^2,k_2^2)  =   (2\pi)^4 \int dm^2  
 \left\{ \left[ \frac{\rho^\pm_{s}(m^2)}{ m^2 - s - i\epsilon } - 
  \frac{\rho_{s,R}^\pm(m^2)}{ m^2 - i\epsilon}  \right]   \pm   (s \to u) 
  \right\}  I_n 
 \frac{1}{( \mu^2 - k_1^2 - i\epsilon) (\mu^2 - k_2^2 - i\epsilon)} \ , 
\label{Apm}   
\end{equation}
\end{widetext} 
where in Eq.~(\ref{Apm}), $I_n = (\mu^2)^nd^n/d(\mu^2)^n$. It should be 
noted that as long as  $s$ and $u$ channel  spectral functions are identical,  
subtractions are really only necessary for $A^+$ while they cancel in 
$A^-$. We are now in position to evaluate the two diagrams (direct and 
crossed) that contribute to the hadronic tensor $W$ as shown in Fig.~\ref{W}. 

For the symmetric part the leading contribution in the Bjorken limit is given 
by 
 
\begin{widetext} 
\begin{eqnarray}
&& T_+^{\mu\nu} =   -\delta_{\lambda'\lambda}
  e_q^2 \int \frac{d^4k dm^2}{i} \left[   \frac{\rho_s^+(m^2)}
  { \left( \frac{p+p'}{2} - k\right)^2 - m^2 + i\epsilon} 
 - \frac{\rho_s^+(m^2)}{-m^2 + i\epsilon} 
 + (s\to u, k\to -k) \right]  \nonumber \\
& \times &  I_n \left[ \frac{1}{ \left[ ( \frac{\Delta}{2} - k)^2 - \mu^2  
  + i\epsilon \right] \left[ (\frac{\Delta}{2} + k)^2 -\mu^2  
  + i\epsilon \right] }  \right] 
  \nonumber \\
& \times &  \left[ 
  \frac{ \left( k + \frac{q+q'}{2} \right)^\mu k^\nu 
  + (\mu \leftrightarrow \nu) - g^{\mu\nu} \left( k + \frac{q+q'}{2} \right) 
  \cdot  k } {\left( \frac{q+q'}{2} + k \right)^2 + i\epsilon} -  
  \frac{  \left( -k + \frac{q+q'}{2} \right)^\mu k^\nu + 
  (\mu \leftrightarrow \nu) - g^{\mu\nu} \left( -k + \frac{q+q'}{2} \right) 
  \cdot  k }
  {\left( \frac{q+q'}{2} - k \right)^2 + i\epsilon} \right].
\label{Tplus}  
\end{eqnarray}
 \end{widetext} 
Similarly the leading contribution to the antisymmetric part can be written, 

\begin{widetext} 
\begin{eqnarray}
&& T_-^{\mu\nu} =   -\tau^3_{\lambda'\lambda} e_q^2 \int \frac{d^4k dm^2}{i} 
  \left[  \frac{\rho_s^-(m^2) }{ \left( \frac{p+p'}{2} - k\right)^2 - m^2 
  + i\epsilon} - \frac{\rho_s^-(m^2)}{-m^2 + i\epsilon} - 
  (s \to u, k\to -k) \right]  \nonumber \\
& \times &   I_n \left[ \frac{-i\epsilon^{\mu\rho\nu\eta} } 
  { \left[ ( \frac{\Delta}{2} - k)^2 - \mu^2  + i\epsilon \right] 
  \left[ (\frac{\Delta}{2} + k)^2 -\mu^2 + i\epsilon  \right]}  \right] 
 \left[ 
  \frac{ \left(k + \frac{q+q'}{2}\right)_\rho k_\eta } 
  {\left( \frac{q+q'}{2} + k \right)^2 + i\epsilon} + 
  \frac{ \left(-k + \frac{q+q'}{2}\right)_\rho k_\eta } 
  {\left( \frac{q+q'}{2} - k \right)^2 + i\epsilon} \right].
\label{Tminus}  
\end{eqnarray}
 \end{widetext} 
To obtain an expression in terms of structure functions or generalized 
parton amplitudes, one applies a collinear factorization to the quark 
propagator in the last square bracket in Eqs.~(\ref{Tplus}) and 
(\ref{Tminus}). We will first consider the diagonal 
case, $q = q'$. In this case $T$ is the analog of the hadronic amplitude for 
forward virtual Compton scattering, whose imaginary part is proportional to 
the DIS cross-section.

\subsection{The DIS Reaction $\gamma^* p \to \gamma^* p$\label{Sec:vcs}}  

It is convenient to express all momenta in terms of light cone components, 
$a^\mu = (a^+,a^-,a_\perp)$ with  $a^{\pm} = a^0 \pm a^z$, 
$a_\perp = (a^1,a^2)$ and to choose a frame in which, $p = p' = (P^+, 
M^2/P^+,0_\perp)$,   $q=q' = (0,  Q^2/\xb P^+,Q_\perp)$, with 
$-q^2=-q'^2 = Q^2 = Q_\perp^2$. Since the nucleon mass $M$ does not play a 
role in our discussion, for simplicity we will set it to zero. The hard quark 
propagators (the term in the last square bracket in 
Eqs.~(\ref{Tplus},\ref{Tminus})) become  

\begin{equation}
  \frac{1}{ \left( \frac{q+q'}{2} \pm  k\right)^2 +i\epsilon}  
 \to   \frac{\xb}{Q^2}  \frac{1}{(-\xb \pm \frac{k^+}{P^+}  + i\epsilon)} 
\end{equation}
 where following the collinear approximation $k\propto P$ we have ignored 
terms of the order $|k_\perp|/\sqrt{Q^2}$. The leading contribution to the 
numerator comes from the terms that maximally involve the photon momentum;  
the term in the last square bracket in Eq.~(\ref{Tplus}) can be written as

 \begin{equation} 
[\cdots] =   \left[  n^\mu  \wtilp^\nu +  n^\nu \wtilp^\mu - 
  g^{\mu\nu} (n\cdot \wtilp) \right]  \frac{(k^+/P^+)^2} 
  { \left(\frac{k^+}{P^+}\right)^2  -\xb^2  + i\epsilon} 
 \end{equation}
 where we have introduced the vectors, $n^\mu = (0^+,2,0_\perp)$ 
($n\cdot a = a^+$) and $\wtilp^\mu \equiv p^\mu/P^+$. In the next step we 
combine all of the propagators using the Feynman parametrization, and we 
obtain 

\begin{widetext} 
\begin{eqnarray}
T^{\mu\nu}_+ & = &-\delta_{\lambda'\lambda} e_q^2 \left[  n^\mu \wtilp^\nu 
  + n^\nu \wtilp^\mu - g^{\mu\nu} (n\cdot \wtilp) \right] 
   \int \frac{d^4 k}{i}\int dm^2 \int_0^1 dx 
   \left[   \frac{(k^+/P^+)^2}{ \left(\frac{k^+}{P^+}\right)^2  - \xb^2  
  + i\epsilon}  \right]   
\nonumber \\   
 & \times &   \left[  \rho^+_s(m^2) 
I_n \left(  \frac{2(1-x)} 
{ \left[ (k - x p)^2  - x m^2 - (1-x) \mu^2 + i\epsilon \right]^3 }
 - \frac{1}{ -m^2(k^2 - \mu^2  + i\epsilon)^2 }  \right) + 
  (s \to u, k\to -k) \right].
  \nonumber \\
\end{eqnarray}
\end{widetext} 
 Finally we perform the $k^-$ and 
$k_\perp$ integrals  using~\cite{int} 
\begin{equation} 
\int  \frac{dk^- d^2k_\perp}{2i(k^2 + a^2 + i\epsilon)^\alpha} = 
  \pi^2 \frac{\Gamma(\alpha-2)}{\Gamma(\alpha)} \frac{\delta(k^+)} 
  {(a^2 + i\epsilon)^{\alpha-2}} 
\end{equation} 
to obtain, 

\begin{widetext} 
\begin{equation}
T^{\mu\nu}_+  =  \delta_{\lambda'\lambda} e_q^2 \left[  n^\mu \wtilp^\nu +  
  n^\nu \wtilp^\mu - g^{\mu\nu} (n\cdot \wtilp) \right]   
 \int_0^1 dx  \frac{2x} { \xb^2 -x^2 -  i\epsilon} \left[ f_q(x) + 
  {\bar f}_q(x)\right] 
 \end{equation} 
 \end{widetext} 
Here $f_q(x), {\bar f}_q(x)$ are the quark and antiquark structure functions, 
respectively, which are given by 

\begin{widetext} 
\begin{eqnarray}
f_q(x)  &  =  &   \frac{\pi^2}{2} \mu^2  \theta(1-x)   \int dm^2 
  \rho^+_s(m^2) I_{n-1}  \frac{x(1-x)^2}{[xm^2 + (1-x)\mu^2]^2} 
  = f_V(x) + f_R(x) 
\nonumber \\
{\bar f}_q(x) & = & \frac{\pi^2}{2} \mu^2  
 \theta(1-x)   \int dm^2  \rho^+_u(m^2) I_{n-1}  \frac{x(1-x)^2} 
  {[xm^2 + (1-x)\mu^2]^2}  = {\bar f}_R(x)  
\label{sf}  \nonumber \\
\end{eqnarray} 
\end{widetext} 
There is no ``valence'' contribution to the antiquark distribution.  
Increasing $n$ produces more powers of 
$(1-x)$ that soften the propagator, form factors, and simultaneously the 
$x\to 1$, end-point behavior of the parton distribution functions (PDFs), 
as dictated by the Drell-Yan-West relation \cite{Drell:1969km}. The valence 
part of the spectral function vanishes in the limit of 
large-$m^2$, which implies that the valence structure functions are 
proportional to $x$ as $x \to 0$. The 
low-$x$ behavior originating from the Regge part of the spectral function 
is given by 
\begin{widetext} 
\begin{equation}
f_R(x)=(\mu^2)^{1-\alpha_s^+} \frac{x \pi^2\beta_s^+}{2} I_{n-1}\int_0^\infty
\frac{dm^2 (m^2)^{\alpha^+_s}}{(xm^2+\mu^2)^2}   \to   (\mu^2)^{1-\alpha^+_s} 
  I_{n-1} \frac{\pi^2 \beta^+_s }{ 2(\mu^2)^{1-\alpha^+_s}}  
    \left[ \frac{\pi \alpha^+_s} {\sin\pi\alpha^+_s}  
  \frac{1}{x^{\alpha^+_s}} \right] 
   \equiv  \frac{\gamma_{\alpha^+_s}} {x^{\alpha^+_s}}
\end{equation}  
\end{widetext}
and for the antiquark distribution ${\bar f}_q(x)$ one 
needs to replace $s \to u$.  As expected the 
small-$x$ behavior of the structure function is determined by the leading 
high-energy behavior of the parton-nucleon amplitude.

A similar analysis for the antisymmetric part, $T^{\mu\nu}_-$, gives 
\begin{equation} 
T^{\mu\nu}_- =  i e_q^2 \epUD \tau^3_{\lambda'\lambda}  
  \int_0^1 \frac{2\xb} {\xb^2 - x^2 - i\epsilon}  \left[ \Delta f_q(x) + 
  \Delta {\bar f}_q(x) \right] \ \ ,
\end{equation} 
where 

\begin{widetext} 
\begin{eqnarray}
\Delta f_q(x)  &  =  &  \frac{\pi^2}{2}  \mu^2  \theta(1-x)   
  \int dm^2 \rho^-_s(m^2) I_{n-1} \frac{x(1-x)^2} {[xm^2 + (1-x)\mu^2]^2}  
  = \Delta f_V(x) + \Delta f_R(x) 
\nonumber \\
\Delta {\bar f}_q(x) & = & \frac{\pi^2}{2}  \mu^2  
 \theta(1-x)   \int dm^2  \rho^-_u(m^2) I_{n-1} 
  \frac{x(1-x)^2} {[xm^2 + (1-x)\mu^2]^2} = \Delta {\bar f}_R(x).    
  \nonumber \\
\end{eqnarray} 
\end{widetext} 
Since antiquarks are expected to dominate in the sea region, the valence 
part $\rho^-_{u,V}$  can be neglected in this region.  The low-$x$ behavior 
of the spin dependent structure functions is determined by the Regge part and 
is proportional to $1/x^{\alpha^-_u}$ or $1/x^{\alpha^-_s}$ 
for $\Delta f_q(x)$ or $\Delta {\bar f}_q(x)$, respectively. 
We note that the subtraction terms do not contribute to the  hadronic tensor. 
This is related to the small-$x$ behavior of the structure functions, which 
are integrable over the low-$x$ region since we assume $\alpha < 1$. 
 The hard propagators in the collinear 
approximation do not spoil the convergence of the integrals over low-$x$. It 
is important to realize, however, that this need not be the case in 
general. For example in the scalar model it was shown that the full 
$T^{\mu\nu}$ amplitude has a constant component (independent of $Q^2$ and 
$\xb$), the so-called $J=0$ pole contribution in the language of Regge 
phenomenology. This component 
originates from the seagull coupling of both photons to the quark at the same 
space-time point, as required by QED gauge invariance. This interaction alone 
leads to a divergent contribution of the form $\int_0 dx f_q(x)/x$ (as opposed 
to $\int_0 dx f_q(x)$ found above) which gets regulated as $x\to 0 $ precisely 
by the subtraction term~\cite{BCG,Brodsky:1972vv,Brodsky:1971zh}. Thus in the 
scalar case the subtraction term is essential  for producing a finite Compton 
amplitude. 

From this discussion it should be clear that the convergence of the 
low-$x$ integration may be a special rather than a general feature of 
these amplitudes.  In Sec.~\ref{Sec:DVCS} we show that convergence arises for  
DVCS in a different manner than for DIS.

\subsection{Normalization\label{Sec:Norm}}
 
The structure functions $f_q(x)$ and ${\bar f}_q(x)$ represent probability 
densities for finding a quark or antiquark of a particular flavor $q$ in 
the nucleon and as such need to be normalized to the net number of quarks 
of that flavor in the proton, ({\it e.g.} $n_q=(0,1,2)$ for  $s$, $d$ and 
$u$ quarks in the proton, respectively) 
\begin{equation}
\int_0^1 dx [\fqx - \bfqx] = n_q.
\end{equation}
Below we verify that this is consistent with the normalization of the vector 
current which is also sensitive to quark densities.  The normalization of the 
diagonal matrix element of the electromagnetic current,  $J^+(0) = e_q 
{\bar \psi}(0)\gamma^+ \psi(0)$,  is given by 
\begin{equation}
 \langle p\lambda'| e_q J^+_q(0)   |p\lambda\rangle = e_q u(p,\lambda') \gamma^+ 
  u(p,\lambda)F^q = 2P^+ \delta_{\lambda'\lambda}  e_q F^q .
\label{Jp}
\end{equation}
The factor of $2$ on the {\it r.h.s} of Eq.~(\ref{Jp}) comes from the 
relativistic normalization of states and $F^q$ is the contribution to the 
proton charge from the particular quark flavor. In terms of the parton-nucleon 
amplitudes defined in Eq.~(\ref{Adef}), the vector current matrix element is 
given by 
\begin{widetext} 
\begin{eqnarray}
\langle p\lambda' |  e_qJ^+_q(0)  | p\lambda\rangle & = &  - e_q \int 
  \frac{d^4k}{i(2\pi)^4}  Tr [ \gamma^+ A] 
= - e_q \int \frac{d^4k} {i(2\pi)^4}  Tr [\gamma^+ A^+\frac{\not k}{4}] 
\nonumber \\ 
&  = &  e_q \delta_{\lambda'\lambda}  \int \frac{d^4 k dm^2 }{i} k^+ 
\left[   \frac{\rho_s^+(m^2)}{ \left( p - k\right)^2 - m^2 + i\epsilon} 
 - \frac{\rho_s^+(m^2)}{-m^2 + i\epsilon} - (s\to u, k\to -k) \right] 
 I_n  \frac{1}{  (k^2 - \mu^2  + i\epsilon)^2 }  \nonumber \\
 & = &  2 P^+  \delta_{\lambda'\lambda} e_q \int_0^1 dx [\fqx - \bfqx ] .
\end{eqnarray}
\end{widetext} 
Thus as expected the quark and antiquark structure functions contribute with 
opposite signs. We also note that the subtraction terms do not contribute, 
since for these terms the integrand is antisymmetric in $k^+$.  The 
normalization of the spin dependent structure functions  is related to the 
axial current matrix element $J^+_{5}(0) = {\bar \psi}(0) 
\gamma^+\gamma_5 \psi(0)$  
 \begin{equation}
  \langle p\lambda'|J^+_{5q}(0)|p\lambda\rangle  = {\bar u}(p,\lambda')
  \gamma^+\gamma_5 u(p,\lambda) g_A^q  = 2 P^+ g_A^q \tau^3_{\lambda'\lambda} .
\label{Jq5}
\end{equation} 
In Eq.~(\ref{Jq5}) $g^q_A$ denotes the contribution from a single quark flavor 
to the nucleon axial charge, and in terms of the spin-dependent structure 
functions should be given by 

\begin{equation} 
g_A^q = \int_0^1 dx \left[ \Delta \fqx + \Delta \bfqx \right] .
\end{equation} 
Indeed, expressing the axial current matrix element in terms of the 
parton-nucleon amplitude we obtain 
 \begin{widetext} 
\begin{eqnarray}
\langle p\lambda'| J^+_{5q}(0) | p\lambda \rangle 
  & = &  - \int \frac{d^4k}{i(2\pi)^4 }  Tr [ \gamma^+ \gamma_5 A] 
= - \int \frac{d^4k}{i(2\pi)^4 }  Tr [ \gamma^+ A^- \frac{\ks}{4}] 
\nonumber \\ 
&  = &  \tau^3_{\lambda'\lambda} \int \frac{d^4 k dm^2 }{i} k^+ 
\left[ \frac{\rho_s^-(m^2)}{ \left( p - k\right)^2 - m^2 + i\epsilon } 
 - \frac{\rho_s^-(m^2)} {-m^2 + i\epsilon } 
  - (s\to u, k\to -k) \right] 
 I_n  \frac{1}{ (k^2 - \mu^2  + i\epsilon)^2 }  \nonumber \\ 
 & =& 2P^+  \tau^3_{\lambda'\lambda} \int_0^1 dx \left[ \Delta \fqx + 
  \Delta \bfqx \right] .
\end{eqnarray}
\end{widetext} 
In the following Section we will consider the collinear approximation for the 
DVCS amplitude. 
 
\subsection{The DVCS Reaction $\gamma^* p \to \gamma p$\label{Sec:DVCS}}

When $\Delta \ne 0$ it is convenient to choose a frame with the following 
momentum coordinates \cite{Brodsky:2000xy} 
(where again we ignore the nucleon mass), 
$p = [P^+,0,0_\perp] $, $p' = [(1-\zeta) P^+, \Delta^2_\perp/(1-\zeta)P^+, 
\Delta_\perp]$, 
$q=[0, (Q_\perp - \Delta_\perp)^2/\zeta P^+ + \Delta_\perp^2/(1-\zeta)P^+,
Q_\perp]$, $q' = [\zeta P^+, (Q_\perp - \Delta_\perp)/\zeta P^+, Q_\perp - 
\Delta_\perp]$. In the Bjorken limit, at small momentum transfer, $\zeta = 
\xb + O(-t/Q^2)$ and $-t = \Delta_\perp^2/(1-\zeta)$. Since we are interested 
in the small-$t$ region we also set $\Delta_\perp = 0$ which also implies 
$\Delta^2 =0$ ($\Delta \to  
[-\zeta P^+, 0,0_\perp]$).  To facilitate comparison 
with standard formulas it is convenient to shift the integration variable in 
Eqs.~(\ref{Tplus}),~(\ref{Tminus}) from $k$ to $\tilde k \equiv  k + 
\Delta/2$.  In the collinear approximation, the hard propagators become  
 \begin{eqnarray}  
& &  \frac{1} { \left( \frac{q+q'}{2} + \wtilk - \Delta/2 \right)^2 }   
  \pm  \frac{1} {\left( \frac{q+q'}{2} - \wtilk + \Delta/2 \right)^2 } 
   \nonumber \\
  & = & \frac{1}{(q' + \wtilk )^ 2 + i\varepsilon } \pm 
        \frac{1}{(q - \wtilk )^2 + i\varepsilon }  \nonumber \\   
  & = & \frac{\xb} {Q^2} 
  \left[  \frac{1}{ \frac{{\wtilk}^+}{P^+} + i\epsilon } \pm 
  \frac{1}{-\xb - \frac{{\wtilk}^+}{P^+} + i\epsilon } \right] .
 \end{eqnarray}
Next we combine the two soft propagators, 
\begin{equation}
 \frac{1}{ \left[ ( \Delta - \wtilk )^2 - \mu^2  \right] \left[ \wtilk^2 
  -\mu^2\right] } 
 =   \int_0^1 dr \frac{1} {\left[ (\wtilk - r \Delta )^2 - \mu^2 
  + i\epsilon\right]^2 }
\end{equation}
and for $T_+^{\mu\nu}$ we obtain, 
\begin{widetext}
\begin{eqnarray}
& & T^{\mu\nu}_+  =  -\delta_{\lambda'\lambda} e_q^2 \frac{1}{2}\left[  n^\mu 
  \wtilp^\nu +  n^\nu \wtilp^\mu - g^{\mu\nu} (n\cdot \wtilp ) \right] 
\int \frac{d^4 \wtilk  dm^2 }{i} \int_0^1dr \int_0^1 dx 
 \left[ \frac{1} { \frac{\wtilk^+}{P^+}  + i\epsilon } - 
  \frac{1}{-\xb - \frac{\wtilk^+}{P^+} + i\epsilon } \right] 
  \left(\frac{\wtilk^+}{P^+} - \frac{\Delta^+}{2P^+} \right)
\nonumber \\   
& \times & 
  \left[  \rho^+_s(m^2) I_n \left(    
  \frac{2(1-x)} {[ (\wtilk - xp' - (1-x) r\Delta)^2 - xm^2 - (1-x)\mu^2 
  + i\epsilon ]^3 } 
  -  \frac{1}{-m^2[ (\wtilk - r\Delta)^2 - \mu^2 + i\epsilon]^2 } \right)
  \right. \nonumber \\
 & & +   \left.  \rho^+_u(m^2) I_n \left(    
  \frac{2(1-x)} {[ (\wtilk + xp - (1-x) r\Delta)^2 - xm^2 - (1-x)\mu^2 
  + i\epsilon ]^3 } 
 -  \frac{1} {-m^2[ (\wtilk - r\Delta)^2 - \mu^2 + i\epsilon]^2 } \right) 
   \right]  \nonumber \\
 \end{eqnarray}
 \end{widetext} 
and after integrating over $\wtilk^-$ and $\wtilk_\perp$, we obtain a formal
relation that is reminiscent of the standard leading-twist DVCS formula in
terms of GPD's. The hadronic tensors, spectral functions and generalized 
parton distributions here all represent the contribution from a single 
quark flavor; we have not included the quark flavor indices but they are 
implicit. As will be discussed shortly, this expression for DVCS 
 fails to be convergent in the presence of Regge behavior. 
\begin{widetext}
\begin{equation} 
 T^{\mu\nu}_+  =  -e_q^2\delta_{\lambda'\lambda} \left[  n^\mu \wtilp^\nu 
  +  n^\nu \wtilp^\mu - g^{\mu\nu} (n\cdot \tilde p) \right]  \left[  
  \int_0^1 dx  H^+(x,\xb) 
 \left(  \frac{1} {x -i\epsilon} + \frac{1} {x - \xb + i\epsilon } \right) + 
  \frac{ f_0(x) +{\bar f}_0(x) }{x}  \int_0^1 dr \frac{(1-2r)^2} {2r(1-r)} 
  \right] .  \\ 
\label{tp0}
\end{equation} 
\end{widetext} 
Here $f_0$ and ${\bar f}_0$ are given by Eq.~(\ref{sf}) without  $(1-x)^2$ in 
the numerator.  Just as in the symmetric case analyzed above in the 
context of DIS, the contribution given by the quark (antiquark) distribution 
$f_q(x)$  ($\bar f_q(x)$) comes from the $s$-channel ($u$-channel) spectral 
function respectively. The $\delta$-function which arises after $\wtilk^-$ 
integration fixes $\wtilk^+/P^+$ in terms of the Feynman parameter-$x$, and leads 
to both positive and negative $\wtilk^+/P^+$.  
  
We immediately note that the last term in Eq.~(\ref{tp0}), which originates 
from the subtractions in the parton-nucleon amplitude needed for the Regge 
term, not only contributes but is in fact singular, since the integral 
diverges for both $r \to 0$ and $r \to 1$ and it has the same sign at both 
limits.  The generalized parton distribution $H^+$ appearing in 
Eq.~(\ref{tp0}) is given by 
\begin{eqnarray}
 H^+(x,\xb)  &  = & (x-\xb/2)\int_0^1 dr \int_0^1 \frac{dy}{y} \nonumber \\
 & \times &   \delta\left[  x -y - (1-y)r\xb \right] 
 \left[ f_q(y)  + {\bar f}_q(y) \right] \label{h}\nonumber \\
\end{eqnarray}
and it is the $C-even$ generalized parton distribution~\cite{Diehl:2003ny}. 
It can easily be checked that $H^+$ satisfies the correct normalization 
conditions  
\begin{equation}
 \int_0^1  dx \frac{H^+(x,\xb)}{1-\xb/2}  = \int_0^1 dx [ f_q(x) + 
  {\bar f}_q(x) ] 
\end{equation}
and 
\begin{equation}
 H^+(x,0) = f_q(x) + {\bar f}_q(x)  .
\end{equation}

Even though the integrals over $x$ and $H^+(x,0)$ are finite, 
$H^+$ is defined by Eq.~(\ref{h}) which is singular.
This can be seen by doing the $y$ integral 
using the $\delta$ function and then changing variables to 
$z = (x - r\xb)/(1 - r\xb)$ and expressing Eq.~(\ref{h}) in the form 
\begin{widetext} 
\begin{equation}
H^+(x,\xb) = \frac{ x - \xb/2}{\xb} \left[ \theta(\xb - x)
  \int_0^x \frac{dz}{z(1-z)} [f_q(z) + {\bar f}_q(z)] 
 + \theta(x-\xb) \int_{\frac{x-\xb}{1-\xb}}^x \frac{dz}{z(1-z)} [f_q(z) 
  + {\bar f}_q(z)]  \right] .
\label{hp}
 \end{equation} 
 \end{widetext} 

First, there is the singularity of  $H^+$ which is of the same type as in 
the Regge subtraction term discussed above. It comes from the lower limit of 
the integral in the term proportional to $\theta(\xb - x)$ in Eq.~(\ref{hp}).  
After integrating over the first hard propagator ($1/[x - i\epsilon]$) in 
Eq.~(\ref{tp0}), the contribution from this singularity to 
the hadronic tensor $T_+^{\mu\nu}$ in the region $x\sim 0$ is given by 
\begin{equation} 
  -\frac{1}{2}\int_0 \frac{dx}{x}  \int_0 \frac{dz}{z} [f_q(z) + \bar f_q(z)] ,
\end{equation}
and after integrating over the second hard propagator $1/[x-\xb + i\epsilon]$, 
in the vicinity $x \sim \xb^-$ gives,
\begin{eqnarray}
 & & +\frac{1}{2} \int^\xb \frac{dx}{x - \xb} \int_0 \frac{dz}{z}  [f_q(z) 
  + \bar f_q(z)] =  \nonumber  \\
  & = &  -\frac{1}{2}  \int^{\xb} \frac{dx}{\xb - x} \int_0 \frac{dz}{z}  
  [f_q(z) + \bar f_q(z)] 
 \end{eqnarray}
 The sum of these two exactly cancels the singularities from the Regge 
subtraction term. 

There are however, residual singularities in the DVCS 
amplitude which originate from the Regge behavior of $H^+$. Consider the 
contribution to the $x$-dependence of $H^+$ from the upper region of 
integration of the term proportional to $\theta(\xb - x)$ in 
Eq.~(\ref{hp}). The low-$x$ Regge behavior of the quark and antiquark 
structure functions is $f_q(x) \sim 1/x^{\alpha^+_s}$ and  
$\bar f_q(x) \sim 1/x^{\alpha^+_u}$, so the quark and antiquark distributions 
give contributions to $H^+$ of the general form   
 \begin{equation}
 H^+(x \sim 0) \sim \frac{1}{2\alpha} \frac{1}{x^{\alpha}} \ . 
 \end{equation}
 The integral over the first hard propagator in Eq.~(\ref{tp0}) thus gives a 
contribution to the DVCS amplitude 
 \begin{equation} 
 \int_0 dx \,H^+(x,\xb) \frac{1}{x - i\epsilon} \sim 
  {\cal O}\left( \frac{1}{\epsilon^\alpha} \right) 
\label{s0} 
 \end{equation} 
 which is divergent for $0<\alpha <1$. 

Similarly, as $x \to \xb^+$  the term in Eq.~(\ref{hp}) for $H^+$ 
proportional to $\theta(x - \xb)$, by virtue of the Regge form for $f_q(z)$ 
and/or $\bar f_q(z) \propto 1/z^\alpha$, is dominated by the lower limit of 
the integral over $z$, leading to 
  \begin{equation} 
 H^+(x \sim \xb^+) \sim \frac{1}{2\alpha} 
  \frac{(1 - \xb)^\alpha} {(x - \xb)^\alpha } 
\label{Hxb}
 \end{equation} 
Using the form of $H^+$ from Eq.~(\ref{Hxb}) in Eq.~(\ref{tp0}), and 
integrating over the second hard propagator, ($1/[x - \xb + i\epsilon]$), 
gives a contribution to the DVCS amplitude of the form 
 \begin{equation} 
 \int_\xb dx \frac{H^+(x,\xb)} {x - \xb + i\epsilon } \sim 
  (1-\xb)^\alpha {\cal O}\left( \frac{1}{\epsilon^\alpha} \right). \label{sb} 
 \end{equation} 
These residual singularities must originate from the collinear approximation 
since after Regge subtraction there is no reason to expect that the 
expression  for $T^{\mu\nu}_+$ in Eq.~(\ref{Tplus}) will be singular. In 
other words, to properly regularize those singularities it will be necessary 
to retain the full momentum dependence of the hard propagators.  

We note that the problem arises from the Regge contribution to the soft part 
of the handbag diagram. The valence spectral functions do not require 
subtraction, thus their contributions to $T^{\mu\nu}_+$ do not have the 
singularity associated with the $(f_{0}(x) + \bar{f}_{0}(x))/x$ 
term in Eq.~(\ref{tp0}). Furthermore valence structure functions vanish at 
small-$x$. As a result, the valence contributions vanish in the regions 
$H^+(x\sim 0)$ and $H^+(x\sim \xb^+)$, so no singularities 
appear of the type given in Eqs.~(\ref{s0}) and (\ref{sb}).

A similar analysis of the antisymmetric contribution yields, 
\begin{widetext}
\begin{eqnarray}
& & T^{\mu\nu}_-  =  -ie^2_q\epUD \tau^3_{\lambda'\lambda} 
 \frac{1}{2}
\int \frac{d^4 \wtilk \,dm^2 }{i} \int_0^1dr \int_0^1 dx 
 \left[ \frac{1}{  \frac{\wtilk^+}{P^+}  + i\epsilon } + 
  \frac{1}{-\xb - \frac{\wtilk^+}{P^+} + i\epsilon } \right] 
\left(\frac{\wtilk^+}{P^+} - \frac{\Delta^+}{2P^+} \right)
\nonumber \\   & \times &  \left[  \rho^+_s(m^2) I_n \left(    
 \frac{2(1-x)} {[ (\wtilk - xp' - (1-x) r\Delta)^2 - xm^2 - (1-x)\mu^2 
  + i\epsilon ]^3 } 
 -  \frac{1}{-m^2[ (\wtilk - r\Delta)^2 - \mu^2 + i\epsilon]^2 } \right) 
 \right. \nonumber \\
 & & -  \left.  \rho^+_u(m^2) I_n \left(    
  \frac{2(1-x)} {[ (\wtilk + xp - (1-x) r\Delta)^2 - xm^2 - (1-x)\mu^2 
  + i\epsilon ]^3 } 
 -  \frac{1}{-m^2[ (\wtilk - r\Delta)^2 - \mu^2 + i\epsilon]^2 } \right)
   \right],  \nonumber \\
 \end{eqnarray}
 \end{widetext} 
\begin{widetext}
\begin{equation} 
 T^{\mu\nu}_-  =  ie_q^2\epUD \tau^3_{\lambda'\lambda}   
   \left[  \int_0^1 dx  \wtilH(x,\xb) 
 \left( \frac{1}{x -i\epsilon} - \frac{1} {x - \xb + i\epsilon} \right) - 
  \frac{ \Delta f_0(x) -\Delta {\bar f}_0(x) }{x}  
  \int_0^1 dr  \frac{(1-2r)} {2r(1-r)}  \right].  
\label{tm0}
\end{equation} 
\end{widetext} 
In this case the infinities arising from the $r=0$ and $r=1$ points in the 
last term in Eq.~(\ref{tm0}) cancel each other, and the contribution from the 
Regge subtraction terms vanishes altogether.  The $\wtilH$ parton 
distribution appearing in Eq.~(\ref{tm0}) is given by the same equation as 
$H^+$ from Eq.~(\ref{hp}) with the replacements $f_q \to \Delta f_q$ 
and $\bar f_q \to  \Delta \bar f_q$,   
\begin{widetext} 
\begin{equation}
\wtilH(x,\xb) = \frac{ x - \xb/2}{\xb} \left[ \theta(\xb - x)
  \int_0^x \frac{dz}{z(1-z)} [\Delta f_q(z) + \Delta  {\bar f}_q(z)] 
 + \theta(x-\xb) \int_{\frac{x-\xb}{1-\xb}}^x \frac{dz}{z(1-z)} 
 [\Delta f_q(z) + \Delta {\bar f}_q(z)]  \right] 
\label{th}
\end{equation} 
\end{widetext} 

In Eq.~(\ref{th}) the Regge parts of $\Delta f_q(z)$ and $\Delta \bar f_q(z)$ 
behave at small $z$ like $1/z^{\alpha^-_s}$ and $1/z^{\alpha^-_u}$, 
respectively.  The singularity of the first integral over $z$ (for $x<\xb$) 
does not contribute to the DVCS amplitude. This is 
because after multiplying by the sum of the two hard propagators the 
contribution of this singularity to $T^{\mu\nu}_-$ in Eq.~(\ref{tm0}) 
becomes proportional to  
\begin{equation}
\int_0^\xb \left(x - \frac{\xb}{2}\right) \left( \frac{1} {x - i\epsilon } 
  - \frac{1} {x - \xb + i\epsilon } \right) dx  = 0. 
\end{equation} 
There are nevertheless the same left-over singularities in the DVCS amplitude 
as in the case of $T^{\mu\nu}_+$.  These originate from the behavior of 
$\wtilH$ near $x\sim 0$ (from the upper limit of the integral in the 
$\theta(\xb - x)$ term), and from the region $x \sim \xb^+$ (from the lower 
limit of the integral in the term proportional to $\theta(x - \xb)$). In the 
region $x\sim 0$ one has the generic behavior $\wtilH \sim 
1/x^\alpha$, so the integral over the first hard propagator is of the 
form, 
\begin{equation}
\int_0^\xb \frac{dx}{x^\alpha} \frac{1}{x - i\epsilon} = {\cal O}\left( 
  \frac{1}{\epsilon^\alpha } \right) .
\end{equation} 
In the region $x\sim x_B$, $\wtilH(x \sim \xb^+) \sim (1 - \xb)^\alpha/
(x - \xb)^\alpha$ and the integral over the second hard propagator becomes 
\begin{equation}
(1-\xb)^\alpha \int_\xb^1  \frac{dx}{(x - \xb)^\alpha} 
  \frac{1}{x - \xb + i\epsilon} 
 \sim (1 - \xb)^\alpha {\cal O}\left( \frac{1}{\epsilon^\alpha} \right) .
\end{equation} 

Even though these singular 
terms contribute to $T^{\mu\nu}_-$ with opposite signs they do not cancel 
because of the extra factor $(1-\xb)^\alpha$.  

\section{DVCS amplitude without collinear approximation\label{Sec:DVCSwo}}
 
In the previous section we noticed that the $C-even$ part of the DVCS amplitude
is singular when evaluated in collinear approximation and expressed in terms 
of the $H^+$ or $\tilde H^+$ GPD's, provided that the parton-nucleon amplitude 
has a high energy behavior typical of hadronic amplitudes, commensurate with 
the Regge type scaling behavior of the form $s^\alpha$ with $0 < \alpha < 1$   
(we also showed that this problem does not arise for the structure
functions).  
From the discussion above it is also clear that the singularity in the DVCS 
amplitude has to do with the collinear approximation to the denominators of 
the hard quark propagator exchanged between photon interaction points.  
Thus in the following we use the collinear approximation for the 
numerators and  keep the full $k$-dependence of the denominators while 
performing the $d^4k$ integral in Eqs.~(\ref{Tplus}) and ~(\ref{Tminus}). 
Then  the Regge part of the spectral function, that is now finite and
dominant at low $t$ in the DVCS amplitude $T^{\mu\nu}_+$ gives,

\begin{widetext}
\begin{eqnarray}
&& T^{\mu\nu}_+ = -\delta_{\lambda'\lambda} e_q^2  \frac{1}{2}   \left[ n^\mu 
\wtilp^\nu + n^\nu \wtilp^\mu - g^{\mu\nu} (n \cdot \tilde p) \right]  
\frac{Q^2} {\xb} \int_0^\infty d\xi \int_0^1 \frac{dx}{x} 
 \int_0^1 dr  \int_0^1 dz 2 \pi^2  (1-x) (1-z)^2
 \nonumber \\
& & 
  \left\{  \frac{\mu^2 \beta^+_s} {(x\mu^2)^{\alpha^+_s} } I_{n-1}    
  \left[  \left( 
 \frac{ \xi^{\alpha^+_s} (1-x)^2[ x(1-\xb)  + \frac{\xb}{2}  - (1-x)(1-z)r\xb 
  - (1-x)z\xb] } { [ \xi + (1-x)(1-z) \mu^2 - (2p\cdot q + q^2) x z 
  (1-x)  - z(1-z)(1-x)^2r q^2 - i\epsilon] ^3 }
  - (x=0) \right) \right.  \right. 
 \nonumber \\
& & \left.  -  \left( \frac{ \xi^{\alpha^+_s} (1-x)^2(x(1-\xb) + \frac{\xb}{2} 
  - (1-x)(1-z)r\xb) }  
   { [ \xi + (1-x)(1-z) \mu^2 + (2 p\cdot q) x z(1-x)  - z(1-z)(1-x)^2 (1-r) 
  q^2 - i\epsilon] ^3 }
 - (x  = 0)  \right)   \right]   \nonumber \\
&&    + 
 \frac{\mu^2 \beta^+_u} {(x\mu^2)^{\alpha^+_u}} I_{n-1} \left[ \left( 
 \frac{ \xi^{\alpha^+_u} (1-x)^2( -x + \frac{\xb}{2}  - (1-x)(1-z)r\xb - 
  (1-x)z\xb) }  
   { [ \xi + (1-x)(1-z) \mu^2 + (2 p \cdot  q) x z(1-x)  - z(1-z)(1-x)^2r q^2 
  - i\epsilon] ^3 }  - (x=0) \right) 
 \right.  \nonumber \\
& &  \left. \left. -  \left( 
  \frac{ \xi^{\alpha^+_u}(1-x)^2(- x + \frac{\xb}{2} - (1-x)(1-z)r\xb) }  
   { [ \xi + (1-x)(1-z) \mu^2 - (2p\cdot q + q^2) x z (1-x)  - z(1-z)(1-x)^2 
  (1-r) q^2 - i\epsilon] ^3 }
 - (x = 0) \right)   \right]  \right\}. 
\label{Tnocoll}  \nonumber \\
\end{eqnarray} 
\end{widetext}

Here following Ref.~\cite{BCG} we changed the $m^2$ variable to $\xi$, with
$m^2 \to \xi/ x$.  The large 
$m^2$ contribution to the integral corresponds to small-$x$ thus we ignore 
$x$ in all terms of the form $(1-x)$, and terms proportional to $x$ in the 
numerator, and we extend the $x$ integral to infinity.  We change the $x$ 
variable so as to bring each 
denominator to the same form as in the subtraction terms (those with $(x=0)$). 
In particular for the term written explicitly  in the second line of 
Eq.~(\ref{Tnocoll}) we replace $x\to x'$ given by, 
\begin{equation}
x = x' \frac{ \xi + (1-z)\mu^2 - z(1-z)r q^2 } {z (2p\cdot q + q^2) } \ \ , 
\end{equation} 
in the third line, 
 \begin{equation}
x = x' \frac{ \xi + (1-z)\mu^2 - z(1-z)(1-r) q^2 } { 2p\cdot q  z } \ \ , 
\end{equation} 
in the fourth line 
\begin{equation}
x = x' \frac{ \xi + (1-z)\mu^2 - z(1-z)r q^2 } { 2p\cdot q z  } \ \ , 
\end{equation} 
and in the fifth line 
\begin{equation}
x = x' \frac{ \xi + (1-z)\mu^2 - z(1-z)(1-r) q^2 } { z (2p\cdot q + q^2) }.
\end{equation} 
We note that since $q^2<0$ and $2p\cdot q + q^2 >0$ these transformations are 
non-singular.  After this change of variables we obtain, 
\begin{widetext}
\begin{eqnarray}
&& T^{\mu\nu}_+ = -\delta_{\lambda'\lambda} e_q^2 \frac{1}{2}   \left[ n^\mu 
  \wtilp^\nu + n^\nu \wtilp^\mu - g^{\mu\nu} (n \cdot \tilde p) \right] 
  Q^2 \int_0^\infty d\xi  \int_0^\infty \frac{dx'}{x'}  
 \int_0^1 dr  \int_0^1 dz 2 \pi^2  (1-z)^2
 \nonumber \\
& & 
  \left\{ \frac{ \mu^2 \beta^+_s} {(x'\mu^2)^{\alpha^+_s} } I_{n-1}  \left[  
  \left[ (2p\cdot q + q^2)z \right]^{\alpha^+_s}
    \frac{  \xi^{\alpha^+_s} [ \frac{1}{2}  - (1-z)r  - z] }  
   { [ \xi + (1-z)\mu^2 + z(1-z) r Q^2]^{3+\alpha^+_s} }  
  \left( \frac{1}{(1 - x' - i\epsilon)^3  } - 1\right)
 \right.  \right.  \nonumber \\
& & \left.  -  \left[ (2p\cdot q) z \right]^{\alpha^+_s} 
  \frac{ \xi^{\alpha^+_s}  [ \frac{1}{2}  - (1-z)r ]  } 
   { [ \xi + (1-z) \mu^2  + z(1-z) (1-r) Q^2 ]^{3+\alpha^+_s} } 
  \left( \frac{1} {(1 + x' - i\epsilon)^3 } - 1 \right)
   \right]   \nonumber \\
&&    + 
 \frac{\mu^2 \beta^+_u} {(x'\mu^2)^{\alpha^+_u} } I_{n-1} \left[ 
  \left[ (2p\cdot q)z \right]^{\alpha^+_u}  
 \frac{ \xi^{\alpha^+_u} ( \frac{1}{2}  - (1-z)r - z) } 
   { [ \xi + (1-z) \mu^2 + z(1-z) r Q^2]^{3+\alpha^+_u} } 
  \left( \frac{1}{(1 + x' - i\epsilon)^3 } -1\right)  \right.  
 \nonumber \\
& &  \left. \left.  -   \left[(2p\cdot q + q^2)z\right]^{\alpha^+_u} 
  \frac{ \xi^{\alpha^+_u} ( \frac{1}{2}  - (1-z)r) }
   { [\xi + (1-z) \mu^2  + z(1-z) (1-r) Q^2]^{3+\alpha^+_u} } 
  \left( \frac{1}{(1 - x' - i\epsilon)^3 } -1 \right)  \right]  
  \right\} \label{Tchofv} \nonumber \\
\end{eqnarray} 
\end{widetext}
The $\xi$ integral in Eq.~(\ref{Tchofv}) can be performed analytically 
yielding,
\begin{widetext}
\begin{eqnarray}
&& T^{\mu\nu}_+ = -\delta_{\lambda'\lambda} e_q^2 \frac{1}{2}   \left[ n^\mu 
  \wtilp^\nu + n^\nu \wtilp^\mu - g^{\mu\nu} (n \cdot \tilde p) \right] 
  Q^2   \int_0^\infty dx'  \int_0^1 dr  \int_0^1 dz 2 \pi^2  
\nonumber \\   
   && \left\{ 
  \frac{z^{\alpha^+_s} \beta^+_s (\mu^2)^{1 - \alpha^+_s} } 
  {(1+\alpha^+_s)(2+\alpha^+_s) }  I_{n-1}       
  \frac{ \frac{1}{2}  - (1-z)r  - z } { [ \mu^2 + z r Q^2]^2 } 
   \frac{1} {x'^{1+\alpha^+_s} }   
   \left[  \left[  2p\cdot q + q^2  \right]^{\alpha^+_s}
    \left( \frac{1}{(1 - x' - i\epsilon)^3 } - 1\right)
 +  \left[ 2p\cdot q \right]^{\alpha^+_s}  \left( 
   \frac{1}{(1 + x' - i\epsilon)^3 } - 1 \right)  \right]  \right.
 \nonumber \\
&&    + \left.
 \frac{z^{\alpha^+_u} \beta^+_u (\mu^2)^{1 - \alpha^+_u} }  
  {(1+\alpha^+_u)(2+\alpha^+_u) }   I_{n-1} 
  \frac{ \frac{1}{2} - (1-z)r - z } {[\mu^2 + z r Q^2]^2 } 
  \frac{1}{x'^{1 + \alpha^+_u} }  \left[ 
  \left[ 2p\cdot q\right]^{\alpha^+_u}  
  \left( \frac{1}{(1 + x' - i\epsilon)^3 } -1\right)   +  
  \left[ 2p\cdot q + q^2\right]^{\alpha^+_u} 
  \left( \frac{1}{(1 - x' - i\epsilon)^3 } -1 \right)  \right]
    \right\} 
\label{Tdvcs} \nonumber \\
\end{eqnarray} 
\end{widetext}

The crucial ingredient which leads to the difference between the DVCS amplitude
given in Eq.~(\ref{Tdvcs}) and the DIS case studied in~\cite{BCG}  
is the presence of the $r$ factor in the $z r Q^2$ terms in the denominators.
In the absence of this factor, the integral over $z$ would be 
dominated by the region  $z \sim \mu^2/Q^2$.  In that case the factors of 
$z^\alpha (2p\cdot q + q^2)^\alpha$ and 
$z^\alpha (2p\cdot q)^\alpha$ would become $Q^2$ independent; this would 
produce a $Q^2$-independent expression for $T^{\mu\nu}_+$ as expected from 
scaling. This additional $r$-dependence is of the same type as found in 
Ref.~\cite{Szczepaniak:2006is}.  In that paper, Regge behavior was introduced 
by utilizing a $t$-channel approach, and not through the $s$ or $u$-channel 
formalism as employed here. The factor of $r$ in $z r Q^2$  makes $r$ peak 
around $\mu^2/Q^2$, and this produces an overall $(Q^2)^\alpha$ dependence 
for the DVCS amplitude. In particular we can write the symmetric tensor 
$T^{\mu\nu}_+$ in the form   
 
\begin{widetext}
\begin{equation} 
 T^{\mu\nu}_+ = -\delta_{\lambda'\lambda} e_q^2  \left[ n^\mu \wtilp^\nu 
  + n^\nu \wtilp^\mu - g^{\mu\nu} (n \cdot \wtilp) \right] \left[ 
  \left(  \frac{Q^2} {\xb \mu^2 } \right)^{\alpha^+_s} F^+_s(\xb) + 
 \left( \frac{Q^2} {\xb \mu^2}\right)^{\alpha^+_u} F^+_u(\xb) \right] 
\label{notscaling}
\end{equation}
\end{widetext} 
where in Eq.~(\ref{notscaling}) we have introduced the quantities 
\begin{widetext} 
\begin{equation}
F^+_{s,u}(\xb) \equiv \frac{\pi^2}{2} 
  \frac{(1-\alpha^+_{s,u}) \Gamma(\alpha^+_{s,u}) } 
   {(1+\alpha^+_{s,u})\Gamma(3 + \alpha^+_{s,u}) }  \beta^+_{s,u} 
  \left[ \mu^2  I_{n-1} \frac{1}{\mu^2} \right]  
  \left[  \xi^+_{\alpha^+_{s,u}}  +  ( 1 - \xb)^{\alpha^+_{s,u}} 
  \xi^-_{\alpha^+_{s,u}} \right]
\label{Fdef}  
 \end{equation} 
 \end{widetext} 
 and in Eq.~(\ref{Fdef}) we define 
 \begin{equation} 
 \xi^\pm_\alpha \equiv \int_0^\infty \frac{dx'} {x'^{1+\alpha} } \left[  
  \frac{1}{(1  \pm x' - i\epsilon)^3  }  -1 \right] . 
\end{equation} 
We call the functions $F(x_B)$ introduced in Eq.~(\ref{Fdef}) the "Regge 
Exclusive Amplitudes," since they contain the information from the coupling 
of the relevant Regge trajectories to a particular exclusive process, 
in this case DVCS.  Unfortunately, the loss of factorization in this 
process makes this and analogous functions non-universal, unlike the 
generalized parton distributions or GPDs. However the Regge Exclusive 
Amplitudes do convey information regarding the exponents $\alpha$ of the 
relevant Regge trajectories 
that are indeed 
universal.  These amplitudes also allow a comparison 
between hard exclusive processes and high-energy total cross-sections.
Alternatively one can directly employ the $t-$channel formulation of the 
hard process in terms of a single (or a few) Regge trajectories.

Finally a similar analysis for the antisymmetric amplitude yields a 
form 
\begin{widetext} 
\begin{equation} 
T^{\mu\nu}_- =  i e_q^2 \epUD \tau^3_{\lambda'\lambda} \left[ 
  \left( \frac{Q^2} {\xb \mu^2} \right)^{\alpha^-_s} F^-_s(\xb) + 
  \left( \frac{Q^2} {\xb \mu^2} \right)^{\alpha^-_u} F^-_u(\xb) \right ]
\label{Tantidef}
\end{equation}
\end{widetext} 
where the relevant Regge Exclusive Amplitudes are defined as 
\begin{widetext} 
\begin{equation}
F^-_{s,u}(\xb) \equiv \frac{\pi^2}{2} 
  \frac{(1-\alpha^-_{s,u}) \Gamma(\alpha^-_{s,u}) } 
   {(1+\alpha^-_{s,u})\Gamma(3 + \alpha^-_{s,u}) }  \beta^-_{s,u} 
  \left[ \mu^2  I_{n-1} \frac{1}{\mu^2} \right]  
  \left[ \xi^+_{\alpha^-_{s,u}}  -  ( 1 - \xb)^{\alpha^-_{s,u}} 
  \xi^-_{\alpha^+_{s,u}} \right]. 
\label{Fantidef}
 \end{equation} 
 \end{widetext} 

We note the familiar structure. The finite constants $\xi^{\pm}_\alpha$ encode 
the integration over the hard propagators from the collinear approximation, 
and contribute with a relative factor of $\pm (1-\xb)^\alpha$ to the symmetric 
and antisymmetric DVCS amplitudes, respectively. This is the same factor that  
arises from the singularities of the collinear approximation. The 
regularization of the collinear approximation leads to an increase of the 
hard exclusive amplitude by a factor of $(Q^2/\xb\mu^2)^\alpha$ relative 
to the DIS amplitude. This is the same enhancement factor that was found in 
Ref.~\cite{Szczepaniak:2006is}. In the more general case, when the nucleon 
and/or quark masses are kept finite or more than one scale appears in the 
parton-nucleon amplitude, the single quantity $\mu$ would be replaced by some 
combination of quantities.  The functions $F^+_{s,(u)}$ describe the quark
(antiquark) helicity averaged contribution to the DVCS amplitude.  
Similarly, the functions $F^-_{s,(u)}$ describe the quark 
(antiquark) helicity-dependent contribution to the DVCS amplitude.

\begin{figure}
\includegraphics[width=2.5in,angle=270]{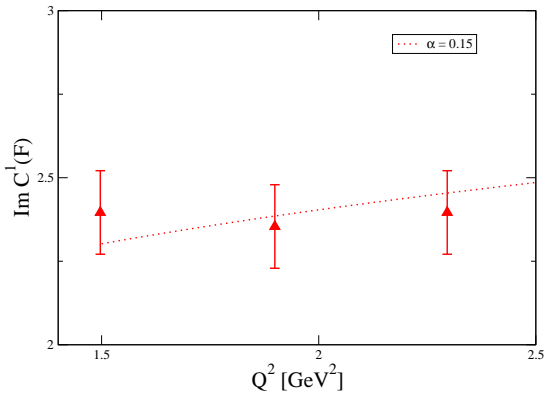}
\caption{\label{datos1} (color online) Comparison with DVCS results 
from Jefferson Lab~\cite{MunozCamacho:2006hx}. The data points represent 
the twist-2 Compton form factor extracted from beam-spin asymmetry 
measurements in $\vec{e}p$ scattering, vs.~$Q^2$. The data have been averaged 
over $t$. The dotted curve is a $(Q^2)^{\alpha}$ fit with $\alpha = 0.15$. 
   }
\end{figure}

As was discussed in Section~\ref{Sec:Intro}, we have carried out a 
preliminary study of photon-induced exclusive processes.  We have shown 
that Regge amplitudes should make significant contributions at large 
values of $x_B$, and not just at small $x_B$.  A major result of our 
formalism is the prediction of scaling violation in these hard exclusive 
processes. At intermediate energies the Bethe-Heitler (BH) amplitude is 
generally substantially larger than DVCS, so DVCS amplitudes must be 
extracted via their interference with the BH term. A group at Hall A in   
Jefferson Laboratory~\cite{MunozCamacho:2006hx} has recently performed 
a test of QCD scaling in spin-dependent $\vec{e}p$ scattering.  They measured 
the beam-spin azimuthal asymmetry \cite{Belitsky:2001ns,Diehl:2003ny}, which 
is proportional to interference between BH and DVCS amplitudes. After 
removing the $Q^2$-dependence associated with the BH term, they extracted 
twist-2 and twist-3 Compton form factors which by QCD scaling 
should be $Q^2$ independent.  In Fig.~\ref{datos1} we plot the twist-2 
Compton form factor $\cal{C^I(F)}$ vs.~$Q^2$; this term has been averaged 
over $t$.  Although the data show very little $Q^2$ dependence, they 
correspond to a limited 
range of $Q^2$ and are also in good agreement with our predicted 
behavior $(Q^2)^{\alpha}$.  In Fig.~\ref{datos1} the dotted line corresponds 
to $(Q^2)^{\alpha}$ with $\alpha = 0.15$.  Because the data points were 
averaged over $t$ it is not obvious what value of $\alpha$ to choose, but over 
this range of $Q^2$ our predicted behavior is in agreement with the 
Hall A points. 
 
\begin{figure}
\includegraphics[width=2.5in,angle=270]{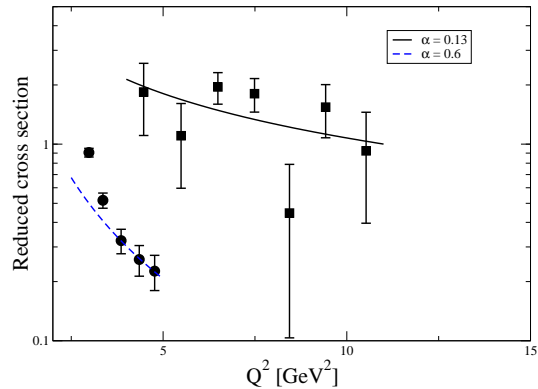}
\caption{\label{datos2} (color online) A simple fit to electroproduction data 
for mesons, $\pi^+$ results from HERMES (squares,~\cite{HERMES-from-jlab}), 
and $\omega$ (circles,~\cite{Morand:2005ex}) results from the CLAS 
Collaboration at Jefferson Lab. In the case of 
$\pi^+$ production the cross section reduced by the photon flux is plotted (in 
arbitrary units).}  
  \end{figure}

In Fig.~\ref{datos2} we compare our predictions with the data on 
exclusive meson electroproduction. Scaling arguments predict that the 
reduced $\pi^+$ cross section should fall off at fixed $\xb$ as $1/Q^2$.  
We predict a behavior $(Q^2)^{2\alpha-1}$ with $0<\alpha  < 1$. Fitting 
$\pi^+$ data from HERMES~\cite{HERMES-from-jlab} in the range  
$0.26 < \xb < 0.8$ gives $\alpha =0.13  \pm 0.1$. Similarly for $\omega$ 
electroproduction cross section from the CLAS collaboration at Jefferson 
Lab~\cite{Morand:2005ex} we find $\alpha =0.6 \pm 0.4$  
for the range $0.52 < \xb < 0.58$. 
  
We see that for both DVCS and exclusive meson electroproduction, not only are 
the data consistent with scaling violations, but the 
additional $Q^2$ dependence is softer than predicted by scaling and in 
agreement with our predicted factor of $(Q^2)^{\alpha}$ with 
$0<\alpha < 1$.  At this point it is difficult to compare the Regge exponents 
$\alpha$ obtained from the fit with total cross-section data, since the 
electroproduction data was taken at different values of $t$. However 
we find this trend encouraging, and we believe that it warrants further 
phenomenological studies. QCD scaling predicts that agreement with 
scaling should become progressively better with increasing $Q^2$.  However we 
have shown that scaling violations should persist regardless of the 
size of $Q^2$. 

\section{Summary and Outlook\label{Sec:Summary}} 
Our formalism in this paper started with the generic hadronic 
tensor for DVCS reactions.  This was then expressed in terms of a 
parton-proton Green's function whose
Regge behavior has been examined in the literature. We have shown how this
makes the standard factorization formula ill-defined due to a collinear
divergence brought about by this Regge behavior; this was demonstrated 
in Eq.~(\ref{tp0}). Appropriate regularization of this divergence, as 
given in Eq.~(\ref{notscaling}), shows that the assumption of $\xb$-scaling 
of DVCS breaks down.  Note that the general analysis of these reactions in 
terms of GPDs assumes $\xb$-scaling.  In our analysis, DVCS and similar 
hard exclusive processes are then characterized by a new set of 
process-dependent Regge Exclusive Amplitudes $F(x_B)$ that are derived 
in Eqs.~(\ref{Fdef}) and (\ref{Fantidef}).  Unlike GPD's, these Regge 
Exclusive Amplitudes are non-universal. One experimental signature of 
this approach is our demonstration that the $Q^2$ dependence of hard exclusive 
amplitudes should differ from scaling predictions, and that one should 
observe a behavior $(Q^2)^\alpha \propto s_{\gamma^* p}^\alpha$
characteristic of hadronic Regge amplitudes. A preliminary examination of 
experimental data on DVCS and hard meson production suggests that the 
data is consistent with the Regge non-scaling $(Q^2)^{\alpha}$ behavior
predicted here.

We argue that the QCD factorization theorems for exclusive processes
\cite{Collins:1996fb,Collins:1998be} should not be applicable to hard 
exclusive processes, at least not in the region of small $t$.  This is due 
to the lack of convergence of the residues of the poles in the $k^-$ plane 
that appear in their derivations. This is a generic feature due to Regge 
behavior.  Thus it appears that the QCD factorization theorems 
for photon-proton exclusive processes necessitate, in addition to a hard scale
 $Q^2$, a sizeable momentum transfer $t$.  This occurs because the Regge 
nature of hadron-hadron amplitudes \cite{Kuti:1971ph} forces a 
corresponding behavior on the parton-nucleon scattering amplitude,which 
generates divergences in the 
Generalized Parton Distribution functions at low $t$.  As $t$ (or $t_{min}$) 
increases the intercept of Regge trajectories become negative, and standard  
collinear factorization then becomes applicable. Thus it appears that 
momentum transfer will become a crucial parameter in these reactions.  At 
small $t$ we predict sizeable effects due to scattering off the meson-cloud 
of the proton, while at large $t$ the dominant effect will become scattering
from the quarks in the "bare" nucleon. The extension of our
results to large $t$ and the detailed interplay with the $J=0$ fixed pole will 
be examined in a future publication \cite{BLS}.

\section{Acknowledgments}
We would like to thank  S.~Brodsky, P.~Hoyer, J-M.Laget, P.~Kroll, D.~Muller, 
A.~Radyushkin and M.~Strikman for useful discussions and comments. 
This work was supported in part by the US Department of Energy grant under 
contract DE-FG0287ER40365, by the US National Science Foundation under grant 
PHY-0555232, and by grants FPA 2004-02602, 2005-02327 (Spain). FJLE thanks 
also the Nuclear Theory Center at Indiana University and the Institute for 
Theoretical Physics at Graz University for their hospitality during the 
preparation of this work.


\begin{thebibliography}{99}
\bibitem{int} For positive (negative)  $k^+$ singularities of the integrand in 
the complex $k^-$ plane are all in the lower (upper) half-plane and the $k^-$ 
integral vanishes. If $k^+ =0$ the integrand is $k^-$ independent and $k^-$ 
integration is divergent. Thus the result has to be proportional to  
$\delta(k^+)$. The coefficient can be determined by integrating over $k^+$ and 
comparing with the covariant result.
\end{thebibliography}
\end{document}